\documentclass[reprint]{revtex4-1}
\usepackage{url} 
\usepackage{lineno}
\usepackage{soul}

\usepackage{amsmath}
\usepackage{amssymb}
\usepackage{graphicx}
\usepackage{bm}

\begin{document}
\title{Turbulence and Transport During Guide-Field Reconnection at the
  Magnetopause}

\author{L.~Price}
\affiliation{IREAP, University of Maryland, College Park MD
  20742-3511, USA}
\author{M.~Swisdak}
\affiliation{IREAP, University of Maryland, College Park MD
  20742-3511, USA}
\author{J.~F.~Drake}
\affiliation{IREAP, University of Maryland, College Park MD
  20742-3511, USA}
\author{D.~B.~Graham}
\affiliation{Swedish Institute of Space Physics, Uppsala, Sweden}



\begin{abstract}

We analyze the development and influence of turbulence in
three-dimensional particle-in-cell simulations of guide-field magnetic
reconnection at the magnetopause with parameters based on observations
of an electron diffusion region by the Magnetospheric Multiscale (MMS)
mission.  Along the separatrices the turbulence is a variant of the
lower hybrid drift instability (LHDI) that produces electric field
fluctuations with amplitudes much greater than the reconnection
electric field.  The turbulence controls the scale length of the
density and current profiles while enabling significant transport
across the magnetopause despite the electrons remaining frozen-in to
the magnetic field.  Near the X-line the electrons are not frozen-in
and the turbulence, which differs from the LHDI, makes a significant
net contribution to the generalized Ohm's law through an anomalous
viscosity.  The characteristics of the turbulence and associated
particle transport are consistent with fluctuation amplitudes in the
MMS observations. However, for this event the simulations suggest that
the MMS spacecraft were not close enough to the core of the electron
diffusion region to identify the region where anomalous viscosity is
important.

\end{abstract}
\maketitle

\section{Introduction}

As the central agent of the Dungey cycle \cite{dungey61a}, magnetic
reconnection controls the interaction between the plasmas of the
magnetosphere and the solar wind.  The necessary change in magnetic
topology occurs at an X-line embedded within a small diffusion region
where kinetic effects become significant enough to make ideal
magnetohydrodynamics an inadequate description of the dynamics, while
the energy stored in the reconnecting fields dissipates on larger
scales, producing flows, heat, and non-thermal particles.  The best
understanding of the details of magnetospheric reconnection springs
from the interplay between {\it in situ} observations and numerical
simulations incorporating the necessary kinetic physics.  The quartet
of spacecraft comprising the Magnetospheric Multiscale (MMS) mission
\cite{burch16c} make temporally and spatially resolved measurements
within diffusion regions but can only sample along individual
trajectories.  Their data complement numerical simulations that
provide synoptic overviews of reconnection but must cope with
computational limitations.

Because of the resources required for fully three-dimensional domains,
many reconnection simulations treat a reduced geometry in which
variations in one direction are ignored.  (In the $LMN$ coordinate
system used here, in which $L$ parallels the direction of the
reconnecting magnetic field and $N$ parallels the inflow direction,
the invariant direction is $M$, which points perpendicular to both $L$
and $N$.  At the equator of the noon-midnight meridian, $L$ points
north-south, $M$ east-west, and $N$ radially.)  This two-dimensional
simplification eliminates fluctuations with non-zero wavenumber $k_M$
and hence inhibits the development of turbulence, which is typically
driven by strong $M$-aligned currents along the magnetopause.
Reconnection in this limit is typically laminar, although
current-driven instabilities along the separatrices can produce
intense parallel electric fields \cite{cattell05a,lapenta11a}.

However, MMS observations of magnetopause reconnection have shown that
turbulence does in fact develop near the X-line, suggesting that
three-dimensional computational domains are necessary
\cite{graham17a,ergun17a}.  Motivated by MMS observations of nearly
anti-parallel (i.e., $B_M \approx 0$) reconnection \cite{burch16a},
simulations demonstrated that the strong gradient in density between
the magnetosheath and magnetosphere triggers a variant of the
lower-hybrid drift instability \cite{price16a,price17a,le17a}.  The
turbulence appeared at both the X-line and along the magnetic
separatrices and had characteristic scale $k\rho_e \sim
(m_eT_e/m_iT_i)^{0.25}$, with $\rho_e$ the electron Larmor radius.  It
relaxed the magnetopause density gradient while producing significant
anomalous resistivity and viscosity in the diffusion region.  The
accompanying fluctuations in the out-of-plane electric field, $E_M$,
had amplitudes much greater than the steady electric field driving
reconnection.

The magnetic configuration during magnetopause reconnection can also
include a significant $B_M$, or guide-field, component and there are
reasons to suspect that this additional component may alter the
conclusions drawn from near-anti-parallel events.  The classic
lower-hybrid drift instability is known to be suppressed by the
presence of magnetic shear, which should be strong when the guide
field is significant \cite{huba82a}.  In addition, magnetization of
the electrons by the guide field, even within the diffusion region
where the reconnecting component of the field vanishes, can affect the
development of anomalous dissipation terms that rely on correlations
between fluctuating quantities.  

On 2015 December 08 at about 11:20 UT MMS passed close to the X-line
of a guide-field ($B_M \approx B_L$) reconnection event at the
magnetopause \cite{burch16b}.  It observed a bifurcated current
system accompanied by significant fluctuations in the current density,
magnetic field, and electric field.  For the perpendicular components
of the electric field those fluctuations were near the local lower
hybrid frequency while the parallel component included higher
frequencies that reached amplitudes of $\approx 30$ mV/m and peaked on
the magnetospheric side of the layer \cite{ergun17a}.  The MMS
observations also reveal that electrons are often frozen-in to the
fluctuations \cite{graham19a}.  Previous simulations based on this
event and two others \cite{le18a} revealed the development of
drift-wave fluctuations as well as noting the accompanying enhancement
of cross-field electron transport and parallel heating. However,
whether the strong turbulence measured at the magnetopause actually
drives transport or produces the necessary dissipation remain open
questions as previous three-dimensional simulations of
reconnection-driven turbulence did not adequately address the impact
of frozen-in electrons on transport
\cite{price16a,price17a,le17a,le18a}.  The turbulence that develops
in the strong density gradient across the magnetopause has
characteristic time scales that are intermediate between the electron
and ion gyrofrequencies. As a consequence, unless electrons resonate
with waves they typically remain magnetized and frozen-in.  Resonance
occurs either through parallel streaming in fluctuations with a finite
$k_\parallel$ with respect to the ambient magnetic field or through
drifts due to the magnetic field gradient \cite{davidson76a}.  It is
known that irreversible transport of plasma density and momentum
requires a resonant interaction between electric fields and particles
\cite{drake81a}. Turbulence with frozen-in electrons greatly reduces
the dissipation described by the generalized Ohm's law.

In this paper we present three-dimensional simulations of reconnection
with initial conditions reflective of the MMS event described in
\cite{burch16b}.  Surprisingly, because of the relaxation of the
cross-magnetopause density gradient and despite the guide field
stabilization, LHDI still develops along the separatrices in a manner
reminiscent of the nearly anti-parallel case.  At the X-line, on the
other hand, the LHDI is stabilized.  Nevertheless, a different
instability develops that produces significant anomalous dissipation.
Finally, we establish that irreversible transport at the magnetopause
can occur, even when the electrons remain frozen-in, due to the strong
vortical motions that effectively create nonlinear fluid resonances.
Section \ref{sims} describes the parameters of the simulation, section
\ref{results} describes the results, while section \ref{discussion}
offers our conclusions.



\section{Simulations}\label{sims}

We perform simulations with the particle-in-cell code {\tt p3d}
\cite{zeiler02a}.  It employs units based on a reference magnetic
field strength $B_0$ and density $n_0$ which then define an Alfv\'en
speed $v_{A0}=\sqrt{B_0^2/4\pi m_in_0}$.  Lengths are normalized to
the ion inertial length $d_i =c/\omega_{pi}$, where $\omega_{pi} =
\sqrt{4\pi n_0 e^2/m_i}$ is the ion plasma frequency, and times to the
ion cyclotron time $\Omega_{i0}^{-1} = m_ic/eB_0$.  Electric fields
and temperatures are normalized to $v_{A0}B_0/c$ and $m_iv_{A0}^2$,
respectively.

The initial conditions closely mimic those observed by MMS during the
diffusion region encounter on 8 December 2015 described in
\cite{burch16b}.  The particle density $n$, reconnecting component of
the magnetic field $B_L$, guide field component $B_M$, and ion
temperature $T_i$ vary as functions of the $N$ coordinate with
hyperbolic tangent profiles of width 1. The asymptotic values of $n$,
$B_L$, $B_M$, and $T_i$ are 0.222, 1.59, -0.659, and 3.19 in the
magnetosphere and 1.00, -1.00, -0.414, and 1.59 in the magnetosheath.
Pressure balance determines the profile of the electron temperature
$T_e$, subject to the constraint that its asymptotic value in the
magnetosphere is 0.664.  (In the asymptotic magnetosheath $T_e$ is
thus 0.159 and, as a consequence, the asymptotic electron pressures
differ by less than 10\%.)  The shear angle between the asymptotic
magnetic fields is $\approx 135^{\circ}$.

Rather than allow reconnection to develop from noise, we apply an
initial perturbation that is uniform in the $M$ direction.  Since such
a perturbation imposes a preferred direction for the development of
the X-line, we have rotated the system so that the $M$ axis bisects
the angle formed by the asymptotic magnetic fields.  Previous work
\cite{swisdak07a,hesse13a,liu15a} suggests that this choice mimics
the direction that the X-line would naturally choose in the absence of
a perturbation.  The initial conditions are not an exact Vlasov
equilibrium, although they are in force balance prior to the
perturbation.  The system adjusts once the simulation begins and
reaches a near-steady-state configuration before the turbulence and
reconnection considered here become important.

We present results from both two-dimensional (in the
$\partial/\partial M=0$ sense) and three-dimensional simulations with
box sizes of $(L_L,L_N) = (40.96,20.48)$ and $(L_L,L_M,L_N) =
(40.96,20.48,20.48)$, respectively.  The boundary conditions are
periodic in all directions.  In order to reduce the computational
expense, but still separate the characteristic scales associated with
the two species, the ion-to-electron mass ratio is chosen to be $100$.
Spatial gridpoints have a separation of $\Delta = 0.016$ while the
smallest physical scale, the Debye length in the magnetosheath,
$\approx 0.04$.  We employ $50$ particles per cell when $n=1$, which
implies $\approx 11$ particles per cell in the low-density
magnetosphere.  To mitigate the resulting noise, our analysis
sometimes includes averages over multiple cells.

The velocity of light in the simulations is $c=15$ so that
$\omega_{pe}/\Omega_{ce}=1.5$ in the asymptotic magnetosheath and
$\approx 0.4$ in the asymptotic magnetosphere while the values derived
from the MMS data are larger ($\approx 40$ and 10, respectively).  As
a result, the ratio of the Debye length to other typical lengthscales
is larger in the simulations, which may tend to suppress very short
wavelength electrostatic instabilities \cite{jaraalmonte14a}.

\section{Results}\label{results}

\begin{figure}
\includegraphics[width=\columnwidth]{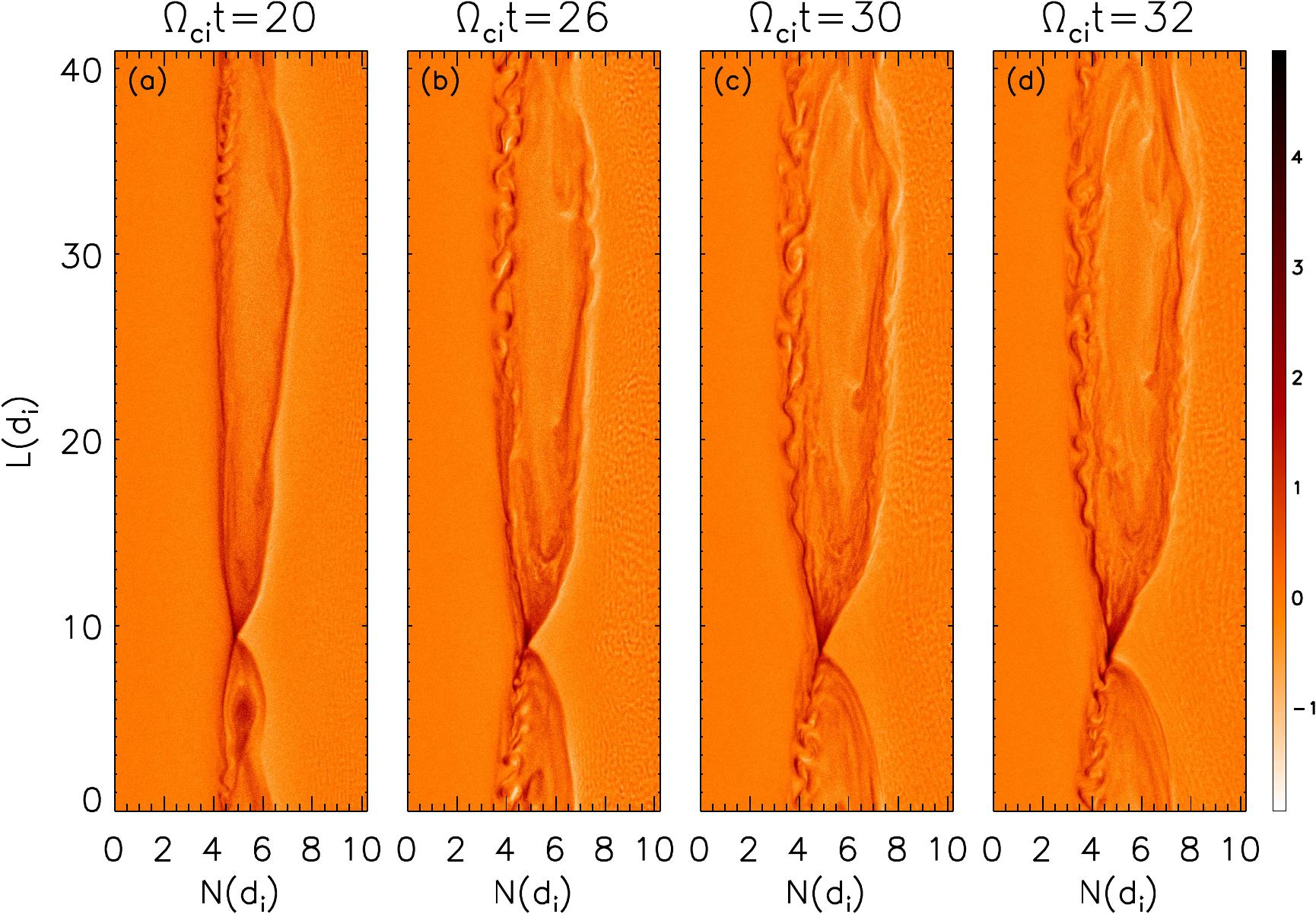}
\caption{\label{jez} Images of $J_{eM}$, the dawn-dusk electron
  current density, in a single $L-N$ plane at four times showing the
  development of turbulence.  The panels share the same normalization,
  which is given by the color bar.}
\end{figure}

Figure \ref{jez} displays images of $J_{eM}$, the dawn-dusk electron
current density, from an $L-N$ slice of the three-dimensional
simulation at four times during which reconnection is occurring at an
approximately uniform rate of 0.11 in normalized units.  In each panel
the magnetosphere (strong field, low density) is to the left and the
magnetosheath (weak field, high density) is to the right.  For
asymmetric configurations such as this the reconnection of equal
amounts of magnetic flux from the two plasmas forces the island to
bulge into the weaker field region.  At the time shown in the first
panel, turbulent features have clearly developed along the downstream
magnetospheric separatrix ($30 \lesssim L \lesssim 40$ and $N \approx
4$).  These arise from the version of the lower-hybrid drift
instability previously seen in simulations of anti-parallel
magnetopause reconnection by others \cite{price16a,price17a,le17a}.
The turbulence expands as the simulation progresses, eventually
appearing along all of both separatrices, with the exception of a
small region near the X-line.  The MMS observations of this event
reveal a double peak in $J_M$, primarily field-aligned, at the closest
approach to the X-line.  Such a bifurcation in $J_M$, also mostly
field-aligned, only occurs in the simulation data at distances
$\lesssim 1 d_i$ ($10 d_e$ for the mass ratio used here) downstream
from the X-point and within the region affected by LHDI.  Close to the
X-line, in contrast, cross-field currents become important.  These
points suggest that the MMS spacecraft did not cross through the
region where, as will be shown below, LHDI is stabilized.  The
two-dimensional companion simulation (not shown) reconnects flux at a
similar rate but remains laminar.

\begin{figure}
\includegraphics[width=\columnwidth]{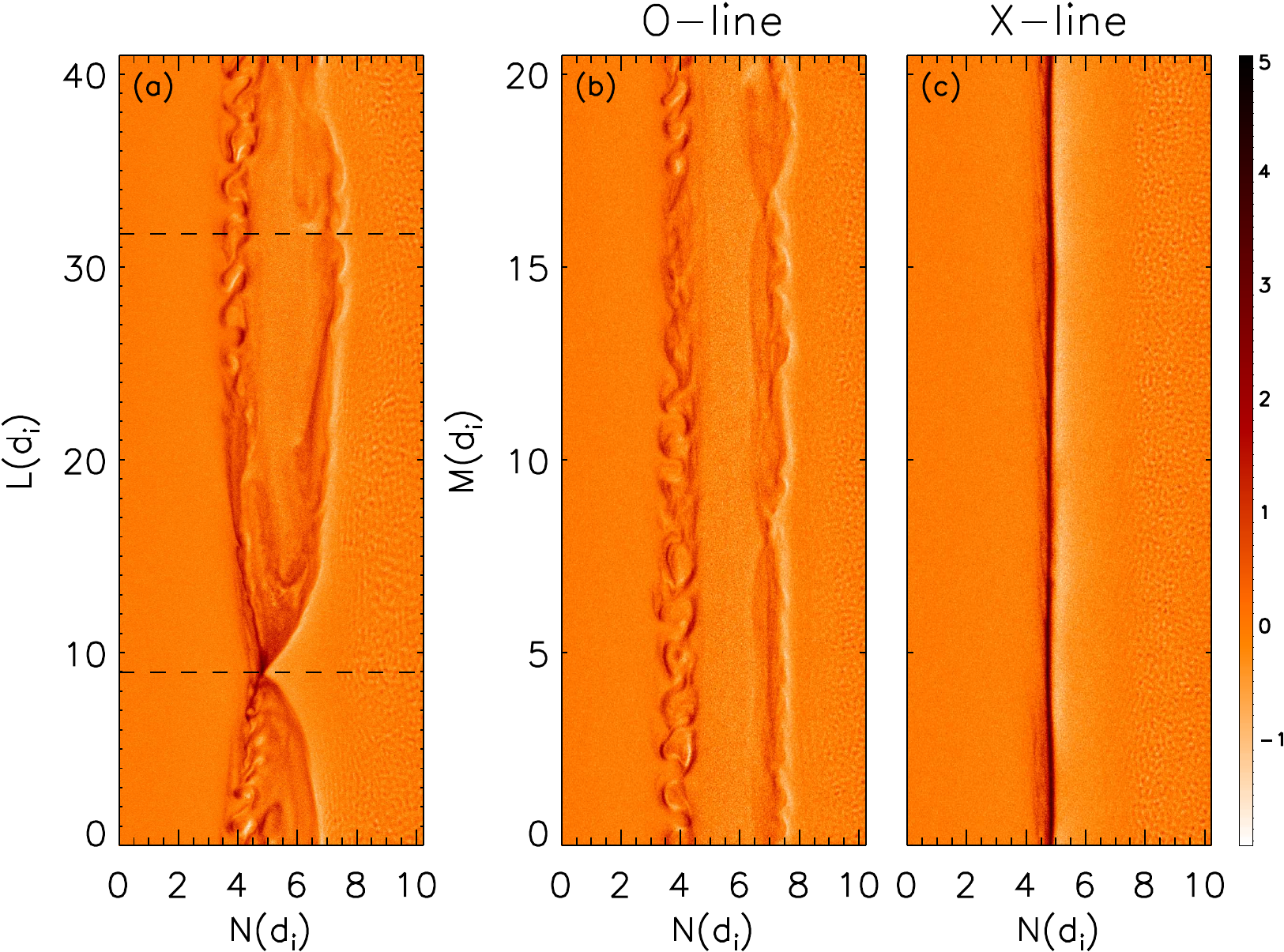}
\caption{\label{jez_x_o} Images of $J_{eM}$ in different planes.
  Panel (a) is identical to panel (b) of Figure \ref{jez} and shows
  the current density in an $L-N$ plane at $t=26$.  Panels (b) and (c)
  show $J_{eM}$ in an $M-N$ plane at the locations indicated by the
  dotted lines in panel (a), the former through the center of the
  island of reconnected flux and the latter through the X-line.}
\end{figure}

Figure \ref{jez_x_o} shows $J_{eM}$ at $t=26$ the approximate time
when the turbulent fluctuations on the separatrices reach their
largest amplitude.  The first panel is identical to panel (b) of
Figure \ref{jez} save for the addition of the two dashed lines giving
the locations of the $M-N$ planes shown in panels (b) and (c).  In the
first, from the center of the island of reconnected flux, the
turbulence at both separatrices is clearly visible although notably
stronger on the magnetospheric side due to, as will be discussed
below, the stronger density gradient there.  The instability is the
same variant of LHDI observed in \cite{price17a} and \cite{le17a} in
simulations of anti-parallel magnetopause reconnection.  In a narrow
current sheet -- one with width less than of order the ion gyroradius
-- theory and simulations suggest a longer-wavelength version of the
classic LHDI develops with the relative drift of electrons and ions in
the $M$ direction supplying the necessary free energy
\cite{winske81a,daughton03a}.  The excited wavenumbers satisfy
$(m_eT_e/m_iT_i)^{0.25} \lesssim k_M\rho_e \lesssim 1$, where $\rho_e$
is the thermal electron Larmor radius.  Using the temperatures from
the time of peak power shown in Figure \ref{jez_x_o}, this can be
written as a condition on the wavelength: $0.24 \lesssim \lambda_M/d_i
\lesssim 1.64$.  The structure in panel (b) has $\lambda_M \approx 0.9
d_i$ and falls within the expected range.  (This agreement should be
qualified by noting that the system includes many strong asymmetries
while most analyses of LHDI assume symmetric Harris-type current
sheets.)


However, a similar cut through the X-line, panel (c), exhibits minimal
turbulence at the wavelengths expected from LHDI.  This is in sharp
contrast to the system with near-anti-parallel reconnection where LHDI
is observed both along the separatrices and on the magnetosphere side
of the X-line.  (Below we will show that there are perturbations at
longer wavelengths unrelated to LHDI.)  The key difference is the
presence of the guide field.  Classic LHDI with $k_{\perp}\rho_e \sim
1$ and $k_{\parallel} = 0$ is known to be stabilized by magnetic shear
at the location of the maximum density gradient \cite{huba82a}.
Stabilization occurs when $k_{\parallel}v_{te}$, where $v_{te}$ is the
electron thermal speed, exceeds $\omega$; for classic LHDI, one can
show that implies stabilization occurs for a magnetic field rotation
of $\Delta \theta \gtrsim \sqrt{m_e/m_i}$.

\begin{figure}
\includegraphics[width=\columnwidth]{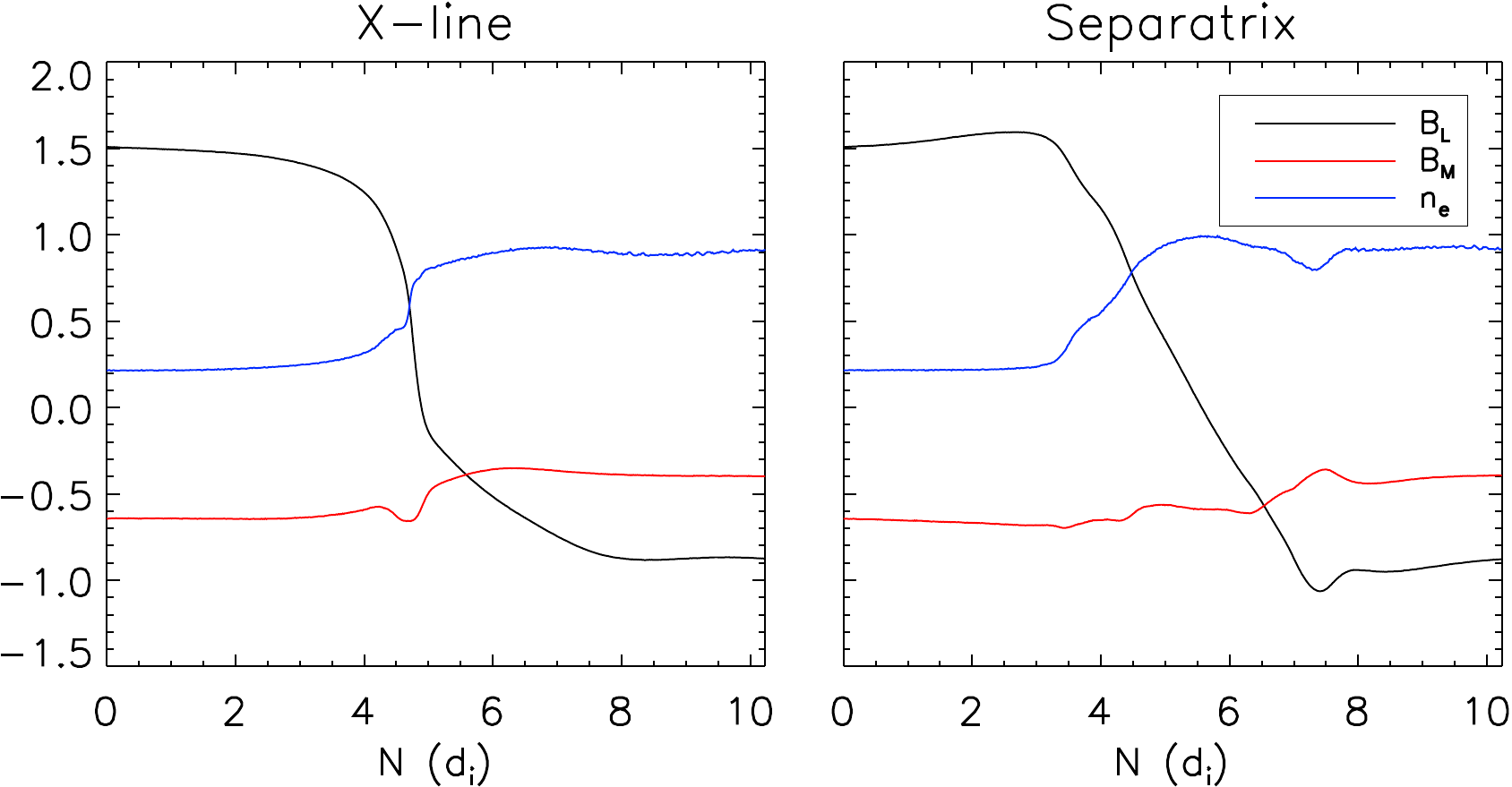}
\caption{\label{shear} Cuts in the $N$ direction through the X-line
  (left) and center of the island (right) at $t=26$ and the locations
  given by the dashed lines in Figure \ref{jez_x_o}a. }
\end{figure}

Figure \ref{shear} shows the profiles of $B_L$, $B_M$, and $n_e$ as
functions of $N$ through both the X-line and the separatrix along the
dashed lines in Figure \ref{jez_x_o}a.  At the X-line the largest
density gradient coincides with a sharp change in $B_L$ (length scale
$\approx 0.5d_i$) and hence a large magnetic shear.  At the
separatrix, in contrast, the rotation in the magnetic field is
significantly weaker (length scale $\approx 1.5 d_i$) and, as a
result, LHDI is not stabilized there.  (Although \cite{huba82a}
considered the classic form of LHDI and not the longer wavelength
variant discussed here the stabilizing influence of magnetic shear is
expected to be similar.)  Figure \ref{shear} also shows the reason for
the relative strengths of the LHDI at the magnetopause and
magnetosheath separatrices: the density gradient at the former
($N\approx 4$) greatly exceeds that at the latter ($N\approx 7$).
While the LHDI is stabilized around the X-line in this guide-field
reconnecting system, we show below that there is a long-wavelength
instability driven by the gradient in the current density that impacts
Ohm's law \cite{che11a}.

\begin{figure}
\includegraphics[width=\columnwidth]{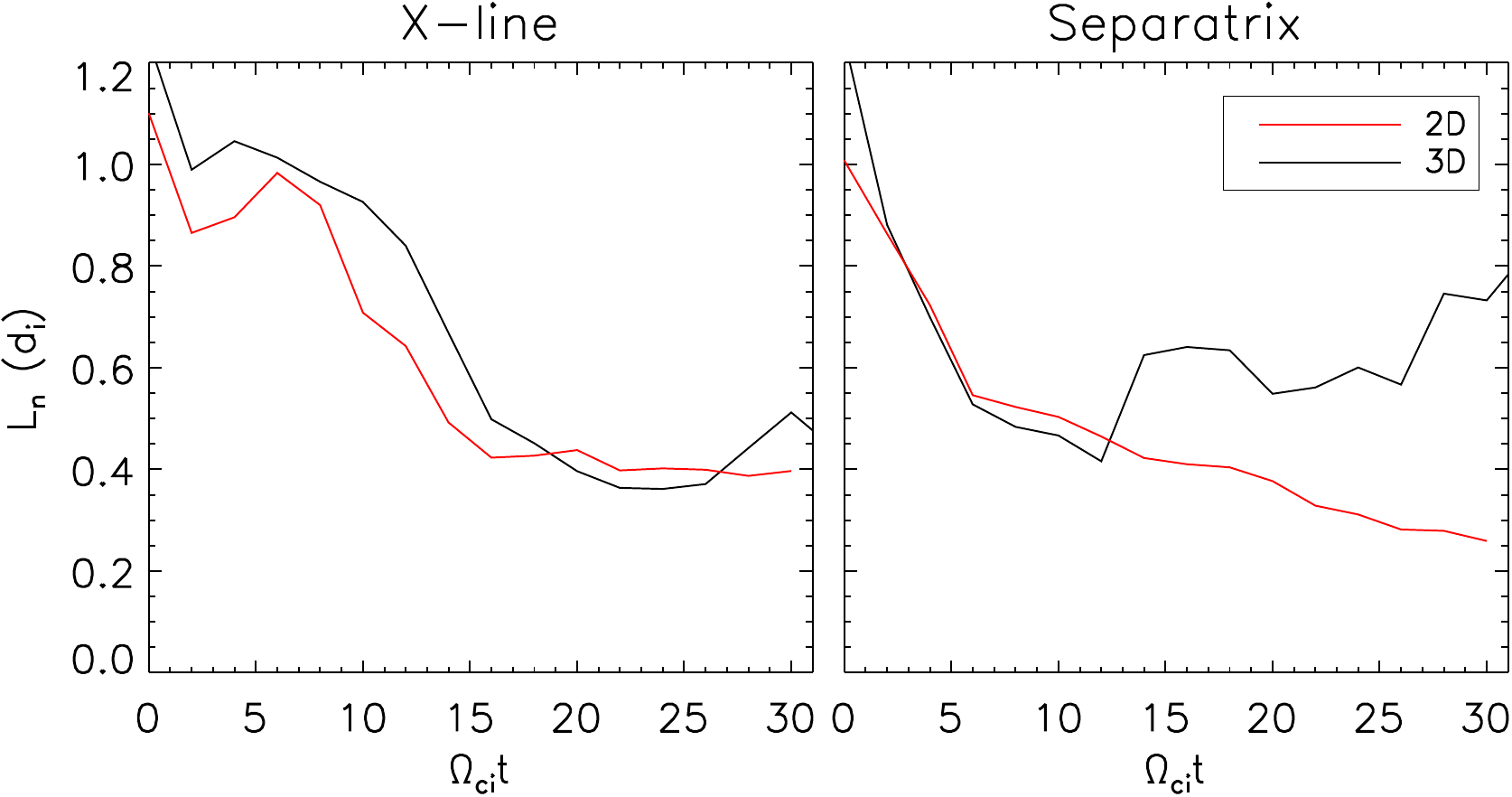}
\caption{\label{denscale} Density scale length $L_n =
  n/|\boldsymbol{\nabla}n|$ measured at the X-line and through the
  downstream separatrix as a function of time for both two-dimensional
  (red) and three-dimensional (black) simulations. }
\end{figure}

As in the anti-parallel reconnecting system, LHDI broadens the current
sheet.  However, due to the magnetic shear stabilization, the
broadening only occurs on the separatrices and not at the X-line.
Figure \ref{denscale} shows the density scale length, $L_n =
n/|\boldsymbol{\nabla}n|$ as a function of time on cuts in the $N$
direction in both the two-dimensional and three-dimensional
simulations.  At the X-line, where the LHDI is suppressed, the
evolution of $L_n$ is remarkably similar in both cases, gradually
shrinking from its initial value until reconnection begins ($t\approx
15$) and then remaining roughly constant. At the separatrix, in
contrast, the two- and three-dimensional simulations only agree until
the amplitude of the LHDI becomes significant, at which point the
three-dimensional current layer broadens and gradually continues to
thicken for the remainder of the run.

Despite the turbulence, magnetic flux reconnects in the
two-dimensional and three-dimensional simulations at essentially the
same rate.  The speed at which this occurs is given by the $M$
component of the electric field, which can be expressed in terms of
other quantities through the generalized Ohm's law, a rewriting of the
electron momentum equation.  However, raw time series of $E_M$ from
the simulation exhibit fluctuations, from both the turbulence and the
noise inherent to PIC simulations, that are much larger in amplitude
than the part of $E_M$ responsible for reconnection.  In order to
quantify the impact of turbulence on large-scale reconnection, we
consider an $M$-averaged version of the generalized Ohm's law that, in
effect, coarse-grains the data.  The averaged Ohm's law is useful if
the scale length of the turbulence is shorter than the length of the
simulation along the $M$ direction. The LHDI develops at short scale
so the averaged Ohm's law yields an appropriate measure of the rate of
reconnection (see panels (b) and (c) of Figure
\ref{jez_x_o}). However, a longer wavelength instability develops at
the X-line (see below). This instability has a wavelength smaller,
though not significantly smaller, than the domain.  Consequently, the
balance of the various terms in the averaged Ohm's law is not as
complete in this location as along the separatrices. A derivation is
given in the Appendix.

\begin{figure}
\includegraphics[width=0.9\columnwidth]{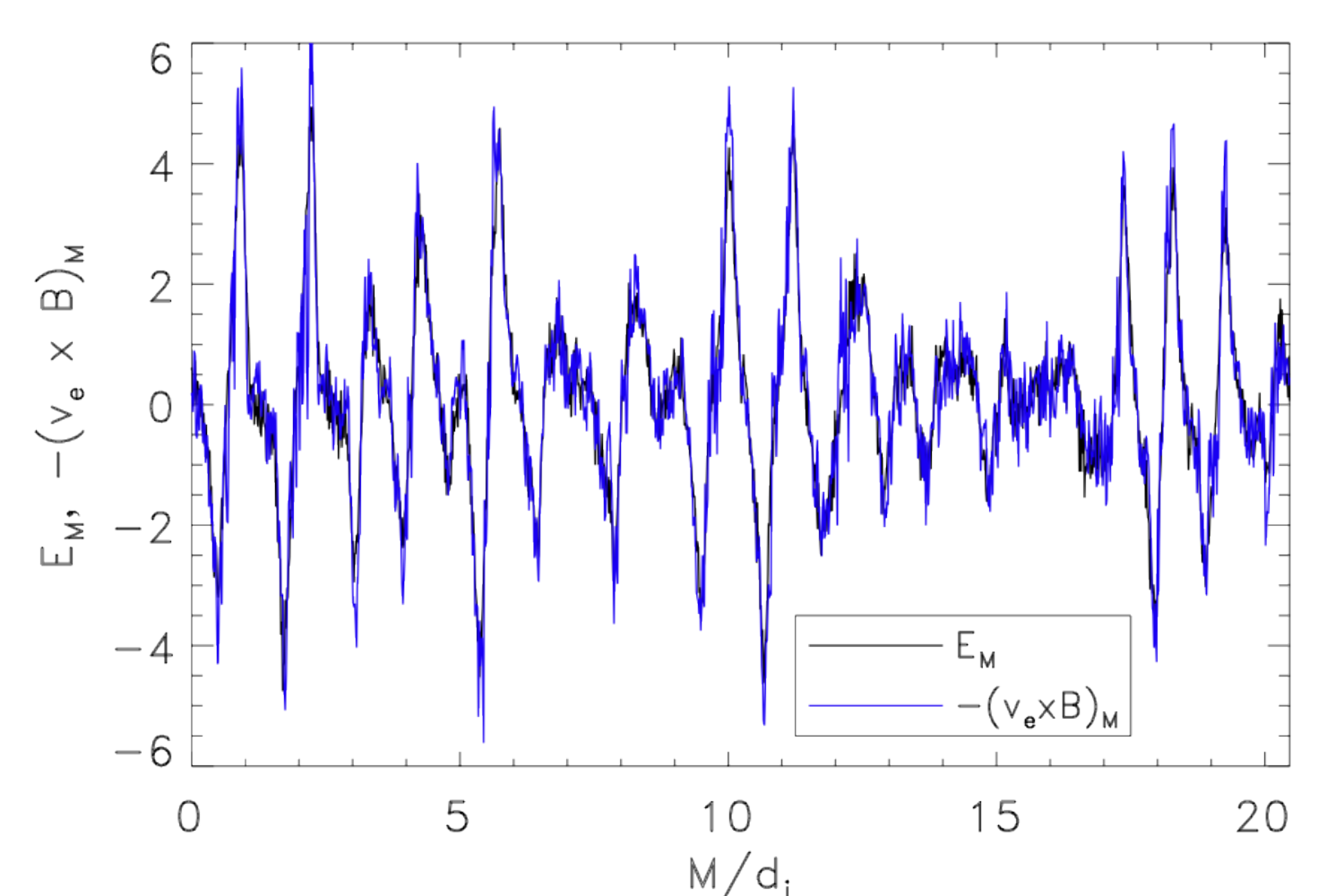}
\caption{\label{frozenin} Cuts in the $M$ direction at the $L$ value
  of the dashed line crossing the separatrix in Figure \ref{jez_x_o}
  and $N\approx 3.9$ of $E_M$ and
  $-(\mathbf{v}_e\boldsymbol{\times}\mathbf{B})_M$.}
\end{figure}

Before discussing the impact of turbulence on how field lines break
during reconnection, we digress briefly on the electron frozen-in
condition in the presence of turbulence. The momentum equation for
electrons is given by
\begin{equation}\label{momeq}
mn\frac{d\mathbf{v}}{dt}= -en\mathbf{E} - \boldsymbol{\nabla
  \cdot}\mathbb{P} - en(\mathbf{v}\boldsymbol{\times}\mathbf{B})/c. 
\end{equation}
In a laminar system the electric field term and the Lorentz force term
balance everywhere except near the X-line where the Lorentz force term
vanishes and the pressure tensor term typically provides the balance
\cite{hesse99a}. Stated another way, the electrons remain frozen-in
everywhere except the X-line. In a system with strong turbulence it is
still possible for the electrons to remain frozen-in to the fluid so
that $\tilde{\mathbf{E}}\sim
-(\tilde{\mathbf{v}}\boldsymbol{\times}\mathbf{B})/c$, where
$\tilde{\mathbf{E}}$ and $\tilde{\mathbf{v}}$ represent turbulent
quantities. Under these conditions, the contributions of the
turbulence to Ohm's law are strongly suppressed because the first and
last terms on the right side of equation (\ref{momeq}) cancel.  In
Figure \ref{frozenin} we show cuts in the $M$ direction of $E_M$ and
$-(\mathbf{v}\boldsymbol{\times}\mathbf{B})_M/c$ at the same value of
$L$ as in Figure \ref{jez_x_o} at the peak of the turbulence along the
magnetopause separatrix ($N\approx 3.9$). The conversion from
simulation to MKS units is such that the peaks in $E_M$ correspond to
fluctuations of $\approx 20$ mV/m, which is reasonably consistent with
the MMS observations \cite{ergun17a}.

The electrons remain almost completely frozen-in even though the
turbulence is strong.  Thus, along the separatrix the impact of the
turbulence on the averaged Ohm's law will be significantly reduced.
That electrons are frozen-in to the LHDI turbulence also raises the
question as to how this instability produces the transport necessary
to broaden the density profile as shown in Fig.~\ref{denscale}.
Resonant interactions that break the frozen-in condition of electrons
are required for real, irreversible transport to occur
\cite{drake81a}.  The data from MMS during its magnetopause crossing
also suggest that electrons remained frozen-in, reducing the impact of
the LHDI on Ohm's law \cite{graham19a}. This cancellation was not
adequately explored in earlier magnetopause simulations
\cite{price16a,price17a,le17a}.

The expansion of the quantities in equation (\ref{momeq}) in terms of
mean and fluctuating terms culminates in equation (\ref{wholething})
after the average over the $M$ direction. The equation for the
reconnection electric field take the form
\begin{equation}\label{dummy}
e\langle n_e\rangle \langle E_M\rangle = (\text{laminar terms}) +
(\text{anomalous terms})
\end{equation}
The first expression on the right-hand side includes terms such as
$e\langle n\rangle \langle v_L \rangle \langle B_N \rangle/c$, only
involves average quantities, and describes the bulk behavior of the
plasma.  These terms represent the contributions to the generalized
Ohm's law from the reconnection electric field seen in both two- and
three-dimensional simulations: the Lorentz force, the pressure tensor,
and the fluid inertia.  When turbulence does not exist or is
unimportant these terms will fully balance the left-hand side of
equation \ref{dummy}.

\begin{figure}
\includegraphics[width=0.9\columnwidth]{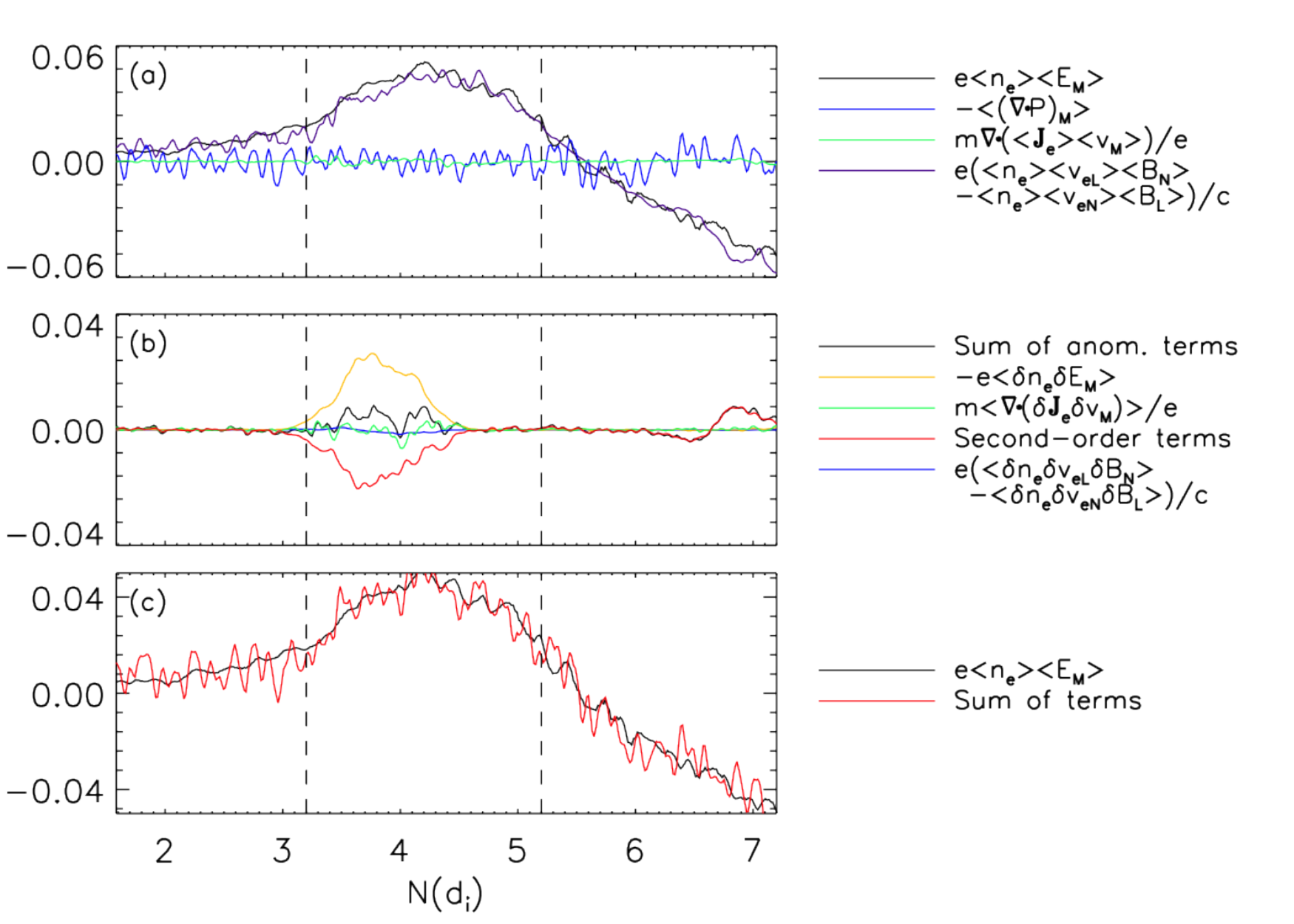}
\caption{\label{ohmplot_sep} Cuts in the $N$ direction through the
  downstream separatrix for the three-dimensional simulation at
  $t=26$. Panel (a): The laminar terms in the $M$ component of the
  $M$-averaged Ohm's law.  Panel (b): The terms in the $M$-averaged
  Ohm's law due to correlations in turbulent fluctuations.  Panel (c):
  The sum of the two sides of equation \ref{ohmeq}.  In each panel the
  vertical lines show the approximate positions of the separatrix
  (left) and the middle of the island (right).}
\end{figure}

\begin{figure}
\includegraphics[width=0.9\columnwidth]{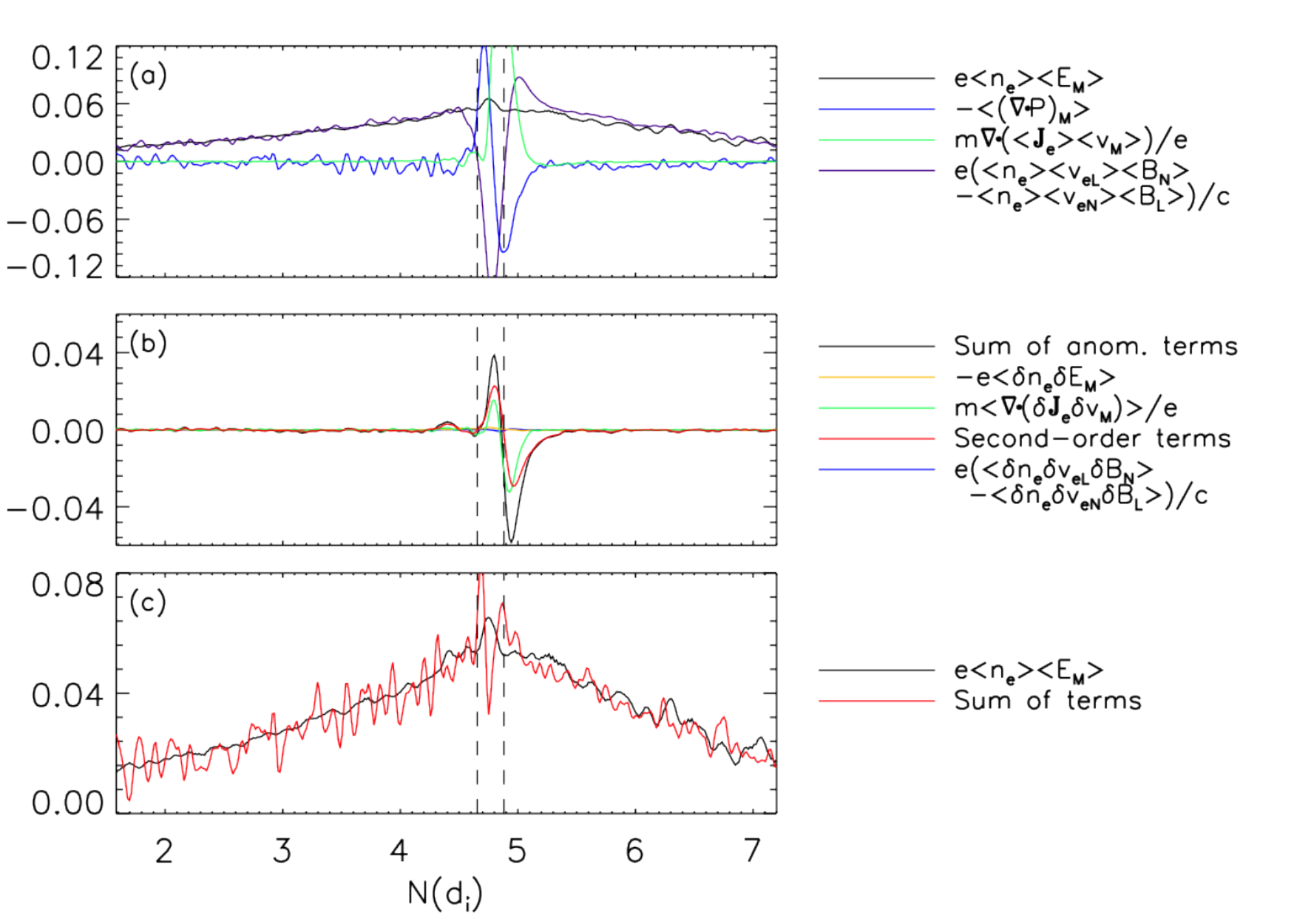}
\caption{\label{ohmplot_x} Cuts in the $N$ direction through the
  X-line for the three-dimensional simulation at $t=26$. Panel (a):
  The laminar terms in the $M$ component of the $M$-averaged Ohm's
  law.  Panel (b): The terms in the $M$-averaged Ohm's law due to
  correlations in turbulent fluctuations.  Panel (c): The sum of the
  two sides of equation \ref{ohmeq}.  In each panel the vertical lines
  show the approximate positions of the stagnation point (left) and
  in-plane magnetic null (right).}
\end{figure}

The second group of terms involves averages of products of fluctuating
components, which can be non-zero in the presence of correlations.
One such term, $\langle \delta n \delta E_M\rangle$, can be
interpreted as arising from an anomalous drag between electrons and
protons \cite{drake03a} while others, including $e\langle \delta n
\delta v_L\rangle \langle B_N \rangle/c$ describe an anomalous
viscosity associated with the turbulent transport of the $M$ component
of the canonical momentum $p_{M}-eA_M/c$ with $B_L=\partial
A_M/\partial N$, where $\mathbf{A}$ is the vector potential
\cite{che11a}. For frozen-in electrons these two terms exactly
cancel.  All of these terms measure the contributions of the
turbulence to reconnection.

In Figure \ref{ohmplot_sep}a we show these terms as a function of $N$
for a cut through the downstream separatrix at the location indicated
by the dashed line in Figure \ref{jez_x_o}.  Panel (a) shows the
laminar terms and demonstrates, as might be expected, that the plasma
is nearly frozen-in with $\mathbf{E} \approx
-(\mathbf{v}\boldsymbol{\times}\mathbf{B})/c$.  The other terms,
representing the contributions of inertia and the pressure tensor, are
negligible.  (The largest component of the pressure tensor divergence,
$\partial P_{MM}/\partial M$, vanishes after the averaging and hence
makes no contribution.)  Panel (b) shows the terms due to the
turbulent fluctuations.  The most significant are the yellow curve
giving the anomalous resistivity and the red curve representing the
six terms due to correlated fluctuations in the elements of
$n\mathbf{v}\boldsymbol{\times}\mathbf{B}$.  The other terms,
represented by the green and blue lines, have negligible effects.  The
black curve shows the sum of all anomalous terms and demonstrates
that, although the largest contributions are similar in amplitude to
the terms shown in panel (a), there is a significant cancellation.
This cancellation confirms that the electrons are basically frozen-in
at this location, which is perhaps not surprising given that the LHDI
wavelength exceeds the electron Larmor radius and the associated
frequency is well below $\omega_{ce}$.  However, seeing that frozen-in
electrons leads to the cancellation of terms in Ohm's law depends on
the decomposition of the generalized Ohm's law leading to equation
\ref{wholething}.  For instance \cite{price17a}'s decomposition used
$\mathbf{J}$ in the Lorentz force term rather than $n$ and
$\mathbf{v}$.  As a result, the frozen-in nature of the electrons and
its impact was obscured both in this paper and others
\cite{le17a}. Finally, in panel (c) the left- and right-hand sides of
equation \ref{wholething} are plotted, showing that the two sides
balance.

Figure \ref{ohmplot_x} shows a similar set of plots as Figure
\ref{ohmplot_sep} from a cut through the X-line along the
dashed line in Figure \ref{jez_x_o}.  The top panel shows that, unlike
the separatrix cut, every term makes a significant contribution to
balancing the reconnection electric field.  Asymptotically the Lorentz
force makes the primary contribution, but between the stagnation and
X-points (left and right dashed lines, respectively) the Lorentz term
reverses sign while the pressure tensor and inertial terms also
contribute.  Panel (b) again shows the anomalous terms.  Unlike at the
separatrices, the anomalous resistivity term, $-e\langle \delta
n_e\delta E_M\rangle$, essentially vanishes, consistent with the
stabilization of LHDI and the laminar cut shown in Figure
\ref{jez_x_o}(c).  The summation of the anomalous terms (black line)
does not vanish, indicating that the electrons are not frozen-in at
this location.  The balance shown in panel (c), while not as precise as
in Figure \ref{ohmplot_sep}(c), is still reasonable.  The deviations
are likely due to the effects of the time-dependent term in Ohm's law,
$\partial \langle J_M\rangle/\partial t$, which is not known to the
same precision as the other terms.

\begin{figure}
\includegraphics[width=0.9\columnwidth]{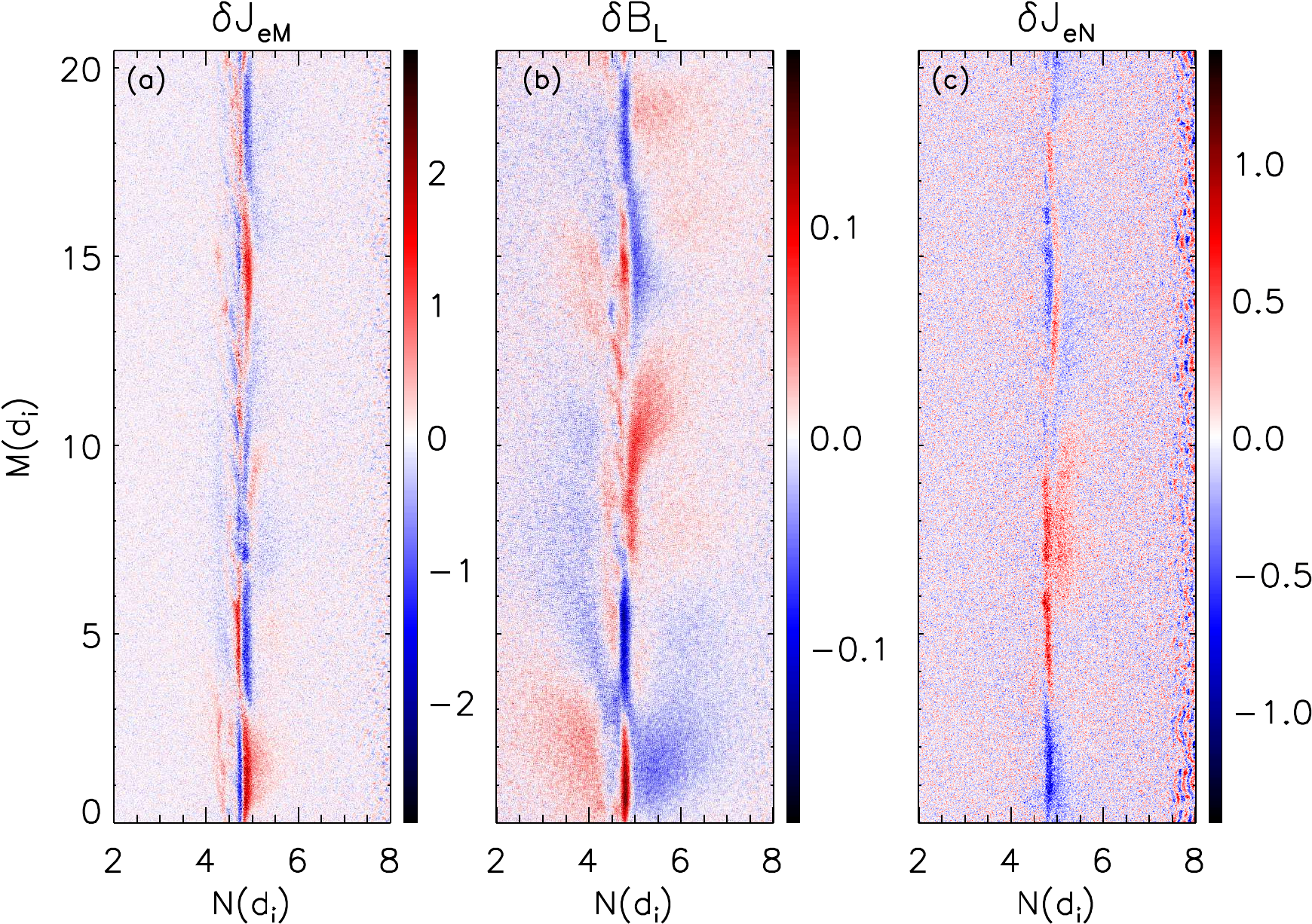}
\caption{\label{flux_x_mn} Images of fluctuating quantities in an
  $M-N$ plane through the X-line at $t=26$.  In each panel white
  represents zero amplitude while red and blue represent positive and
  negative fluctuations, respectively.}
\end{figure}

\begin{figure}
\begin{center}
\includegraphics[width=0.4\columnwidth]{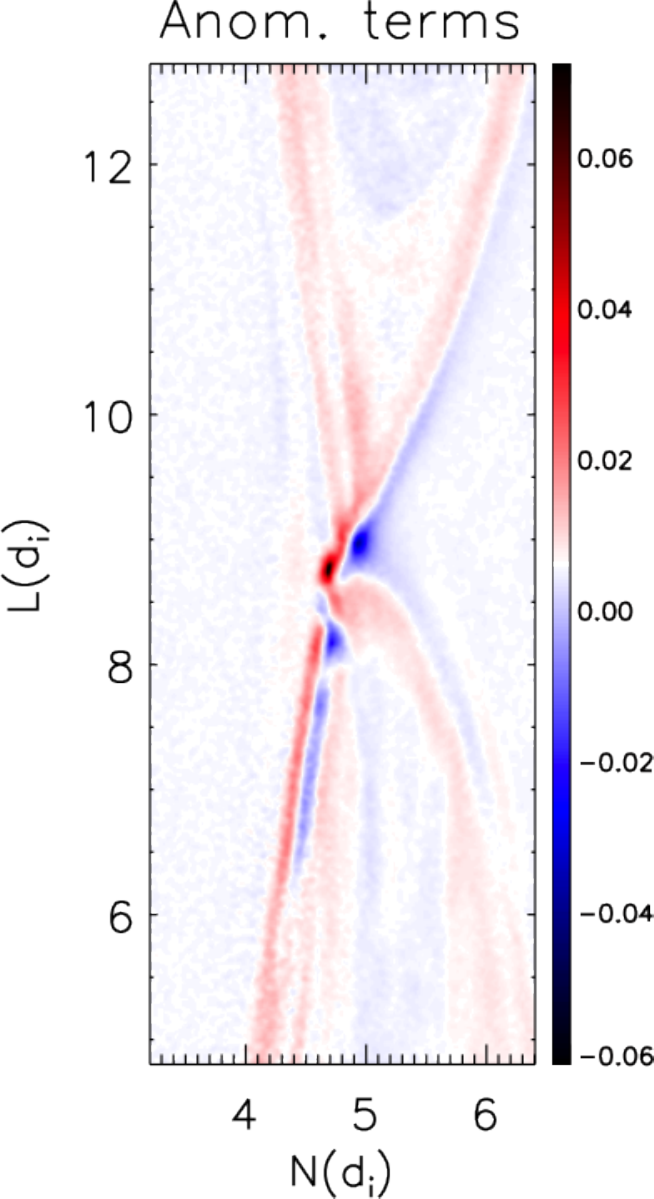}
\caption{\label{visc} Image of the anomalous terms from equation
  \ref{wholething} in a portion of the $L-N$ plane at $t=26$ showing
  its localization near the X-line.}
\end{center}
\end{figure}

What is the source of the anomalous contributions to the generalized
Ohm's law at the X-line in Figure \ref{ohmplot_x}(b) where the
magnetic shear stabilizes LHDI?  Figure \ref{flux_x_mn} displays the
fluctuating parts (i.e., the part remaining after subtracting off the
average in the $M$ direction) of three quantities at the same time and
location as panel (c) of Figure \ref{jez_x_o}: $\delta J_{eM}$,
$\delta B_L$, and $\delta J_{eN}$.  The fluctuations share several
characteristics with the structures reported in \cite{ergun17a} after
analysis of several MMS events.  They observed fluctuations with
$k_{\parallel} \neq 0$ accompanied by parallel electron flows and
strong, high-frequency bursts of $E_{\parallel}$ in regions with
minimal electron pressure gradients.  These characteristics did not
match those usually associated with LHDI, leaving them unsure of the
source of the turbulence.  The mode structure in the simulations is
complex, but approximately two wavelengths fit into the domain.  In an
earlier three-dimensional simulation with half of the length of the
domain in $M$ only a single wavelength of this instability
appeared. The displacement of the electrons in the $N$ direction in
Fig.~\ref{jez_x_o}(c) leads to corresponding increases and decreases
in the local current $\delta J_{eM}$ in Fig.~\ref{jez_x_o}(a),
indicating that the electron motion displaces the ambient current
$J_{eM}$. That the motion is localized in $N$ in the region where the
gradient in $J_{eM}$ is large suggests that this instability is driven
by the current gradient \cite{che11a}.  Figure \ref{visc} shows the
spatial distribution of the contributions of the anomalous terms to
the generalized Ohm's law in a portion of the $L-N$ plane at $t=26$.
The terms are small along the separatrices, consistent with the cuts
shown in Figure \ref{ohmplot_sep}.  Near the X-line, however, there is
a notable increase that is reflected in the cuts seen in Figure
\ref{ohmplot_x}.  The region of enhanced anomalous viscosity extends
$\lesssim 1 d_i$ downstream from the X-line and barely reaches the
region of bifurcated $J_M$.  The guide field also introduces an
asymmetry in the $L$ direction. Close to the X-line, but outside the
electron diffusion region, the anomalous contributions appear to be
more significant southward of the X-line.  Since MMS observations
include a notable current bifurcation, the effects of anomalous
viscosity are not expected to appear in its data.

In the anti-parallel reconnecting system, LHDI drove sufficient
particle transport to limit the density scale length to a hybrid of
the electron and ion Larmor radii both along the separatrices and on
the magnetospheric side of the X-line. We now explore how the
turbulence in the case of a guide field limits the density
profiles. To quantify the transport we write the turbulent particle
flux in the $N$ direction, discarding the component along the magnetic
field $B_N$, as
\begin{equation}
  \Gamma_{N\perp}=\langle \delta n \delta V_{eN,\perp}\rangle,
  \label{nflux}
  \end{equation}
where the average is over the $M$ direction.  Due to the magnetic
shear stabilization around the X-line, the turbulence only controls
the profiles across the separatrices (see Figure \ref{denscale}).  On
the other hand, the electrons remain frozen-in even along the
separatrices so how can the electrons undergo irreversible transport,
which requires a resonant interaction between the electrons and the
turbulence? For turbulence with $k_\parallel=0$, electrons can
resonantly interact with the waves through their
$\boldsymbol{\nabla}B$ drift \cite{davidson76a}. However, because of
the low electron temperatures in the present simulations and at the
actual magnetopause, this resonance is not likely to play a role.  In
linear theory in this limit $\delta n$ is simply given by the electron
convection along the density gradient and is $\pi/2$ out of phase with
$V_{eN,\perp}$. The consequence is that the right side of equation
(\ref{nflux}) is zero unless the amplitude of the wave is changing in
time. The broadening of the density profile in this limit results from
rippling of the density profile and is fully reversible
\cite{drake81a}.

\begin{figure}
\begin{center}
\includegraphics[width=0.4\columnwidth]{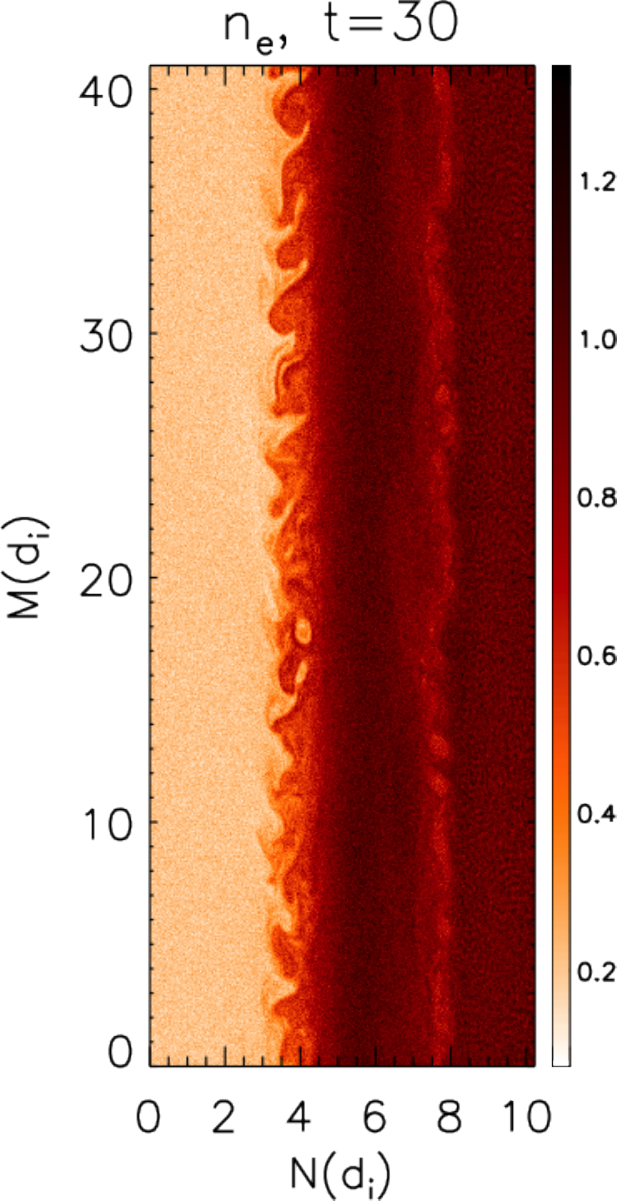}
\caption{\label{dene30} Electron density in the same $M-N$ plane as in
  Figure \ref{jez_x_o}b at $t=30$ showing the development of nonlinear
  structures along the magnetospheric separatrix.}
\end{center}
\end{figure}

\begin{figure}
\includegraphics[width=\columnwidth]{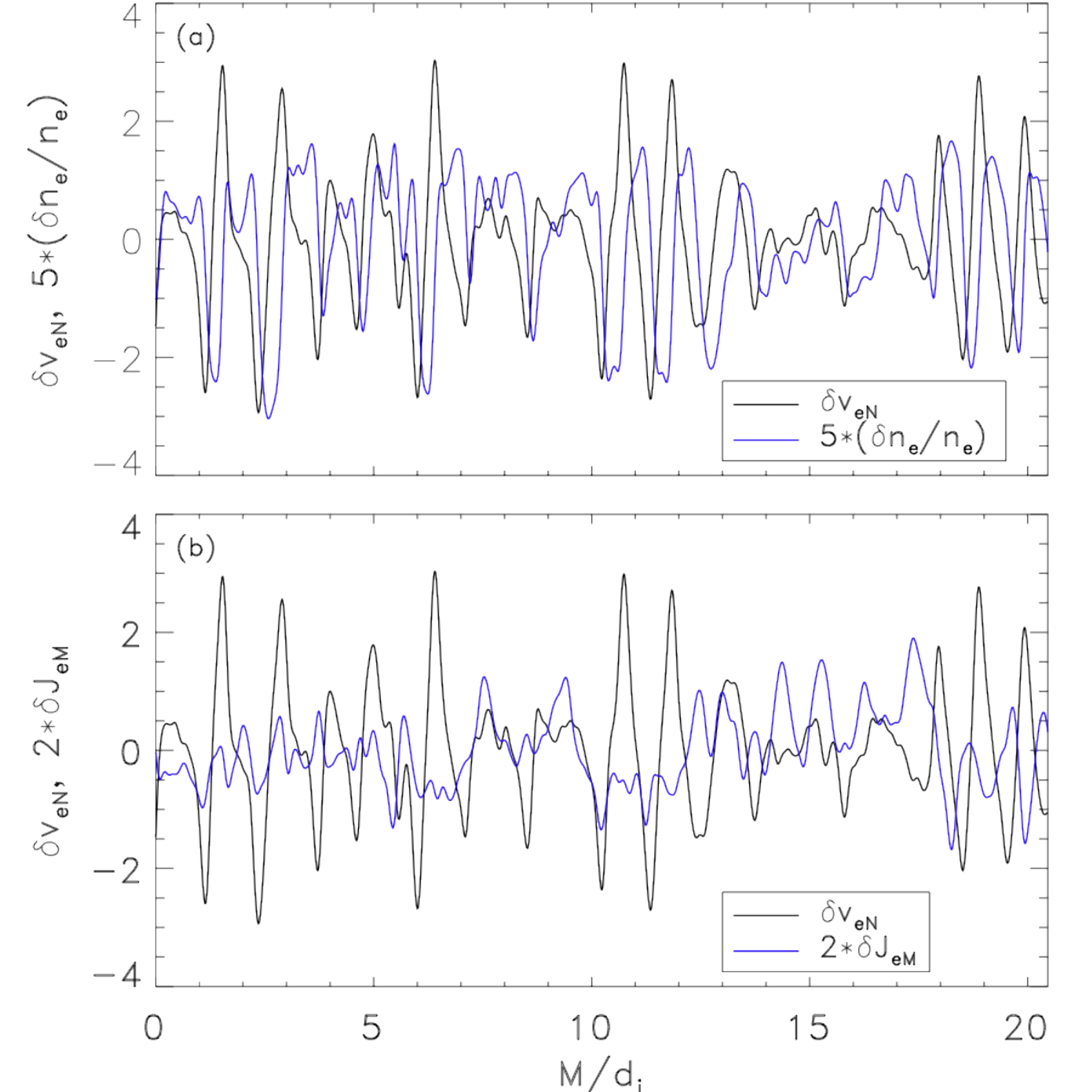}
\caption{\label{turbcuts} Cuts in the $M$ direction at the $L$ value
  of the dashed line crossing the separatrix in Figure \ref{jez_x_o}
  and $N\approx 3.9$ of: (a) $\delta n_e/n_e$ and $\delta V_{eN,\perp}$
  and (b) $\delta V_{eN,\perp}$ and $\delta J_{eM}$.  All quantities
  have been smoothed to remove high-frequency noise.}
\end{figure}

However, irreversible transport can occur via
$\mathbf{E}\times\mathbf{B}$ trapping \cite{kleva84a}, in which the
vortical motion of the fluid is strong enough for the electron orbits
perpendicular to the magnetic field to become chaotic. The result is a
fluid-like rather than a kinetic resonance. The requirement to enter
this regime is for the $M$ component of the
$\mathbf{E}\times\mathbf{B}$ drift to exceed the wave phase speed in
the electron reference frame. In Figure \ref{dene30} we show a cut of
the density in the $M-N$ plane in a cut through the magnetopause
separatrix. At this time the density no longer exhibits the periodic
displacement in the $N$ direction that characterizes the linear LHDI
stability theory. The density has developed a complex turbulent
structure that characterizes true diffusion. Thus, electron diffusion
across the magnetopause is possible even if the electrons remain
frozen-in. Cuts of the individual quantities from the right-hand side
of equation \ref{nflux} that exhibit the phase relationship between
them, are shown in Figure \ref{turbcuts}a.  Because of the nonlinear
nature of the turbulence the phase relation between $\delta n/n$ and
$V_{N,\perp}$ is complex but exhibits regions where the two quantities
are in phase and net transport occurs. Such cuts are closely related
to what spacecraft passing through a turbulent reconnecting separatrix
would observe.

\begin{figure}
\begin{center}
\includegraphics[width=0.8\columnwidth]{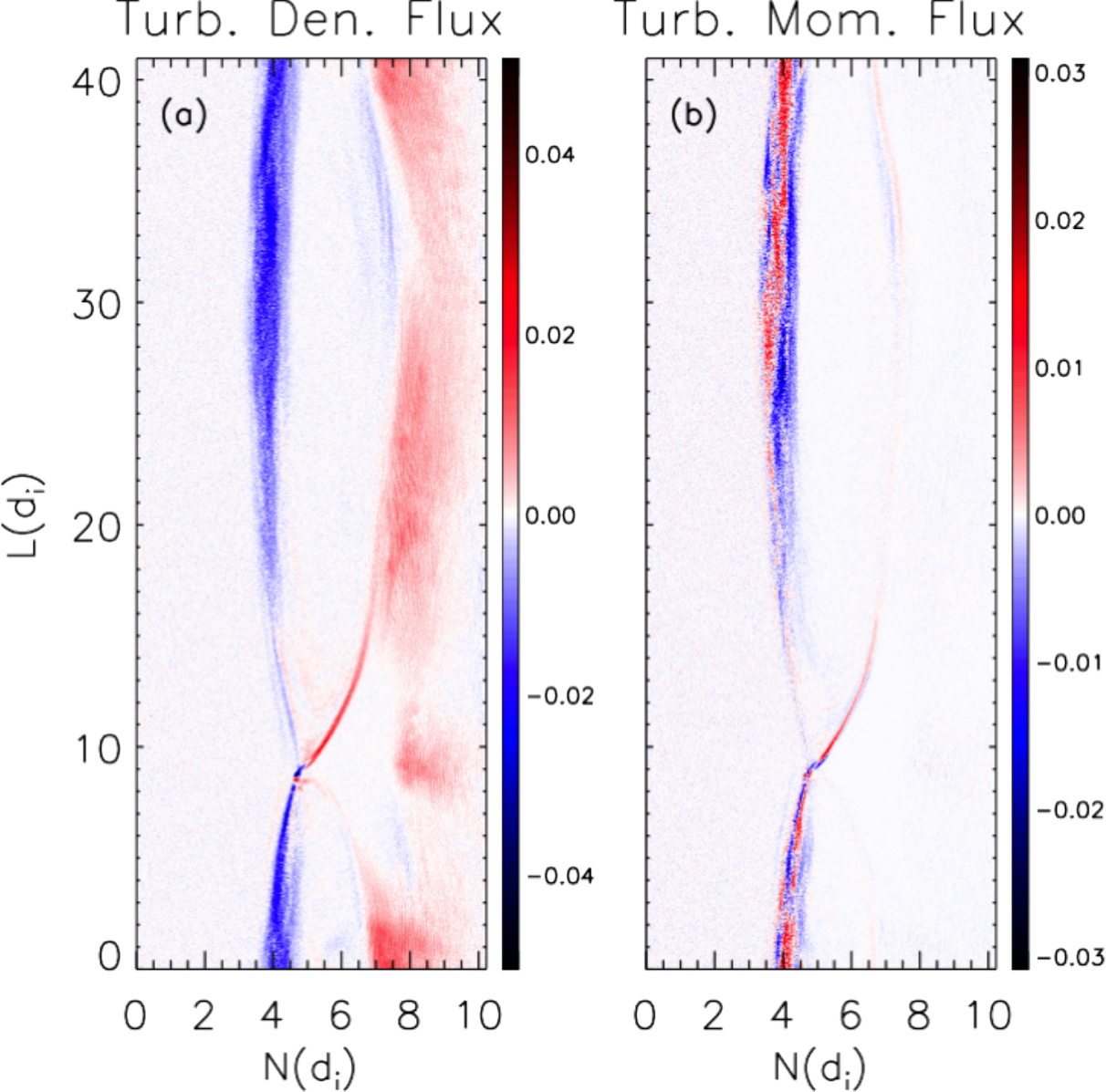}
\caption{\label{turbflux} Images in the $L-N$ plane of (a):
  $\Gamma_{N,\perp}=\langle\delta n \delta V_{eN,\perp})\rangle$ and
  (b): $\langle\delta v_{N\perp} \delta(n v_M)\rangle$ at $t=26$.}
\end{center}
\end{figure}

The flux $\Gamma_{N\perp}$ calculated from the simulation data using
equation \ref{nflux} is shown at $t=26$ in the $L-N$ plane in Figure
\ref{turbflux}a. For the density profiles to reach a
quasi-steady-state as suggested by the data in Figure \ref{denscale},
the fluxes associated with the laminar motion (reconnection inflows
and outflows) both perpendicular and parallel to the ambient field
must balance the diffusive fluxes. This requires $\Gamma_{N\perp}\sim
nV_{in}\sim 0.1nC_{AL}$, where $V_{in}\sim 0.1C_{AL}$ is the
characteristic reconnection inflow speed. The turbulent fluxes in
Figure \ref{turbflux}a are sufficient to balance the laminar flows
associated with reconnection, which is consistent with the
quasi-steady density gradient that develops along the separatrix.

If the electron density can undergo diffusion across the magnetopause
even when the electrons remain frozen-in, it is possible that the $M$
component of the electron momentum (current) can do so as well. In the
case of an anti-parallel magnetic configuration the coupled diffusion
problem has been completed \cite{drake81b}. In the context of the
present simulations with a strong guide field, we are not interested
in solving the full coupled diffusion equations but rather focus on
the diffusion of the electron momentum in the $M$ direction.  Consider
the $M$ component of Ohm's law and assume that the electrons are
completely frozen-in and that the divergence of the electron pressure
can be neglected.  The $M$ component of the momentum equation becomes
\begin{equation}\label{graham}
\frac{\partial (nV_{M})}{\partial t} + \boldsymbol{\nabla
  \cdot}[\mathbf{v}(nV_M)] = 0, 
\end{equation}
which is simply the diffusion equation for the current $nV_M$. Once
the equation is broken into laminar and fluctuating parts, the
transport of momentum is given by the turbulent momentum flux
\begin{equation}\label{currdiff}
\langle\delta V_{N\perp} \delta(n V_M)\rangle,
\end{equation}
where we have retained only the the perpendicular transport of
$nV_M$. The individual components of this equation are again shown in
cuts along the $M$ direction in Figure \ref{turbcuts}b.  The flux of
the out-of-plane current in the $N$ direction normal to $\mathbf{B}$
is shown in Figure \ref{turbflux}b. The flux is strong, particularly
on the downstream magnetospheric separatrix and exhibits an asymmetry
in the north-south direction due to the presence of the guide field.
When normalized by the averaged out-of-plane current density, $\langle
n v_M \rangle$, this flux is larger than the similarly normalized
particle flux shown in Figure \ref{turbflux}b.  The frozen-in nature
of the electrons is a natural consequence of the temporal and spatial
scales associated with the turbulence.  What is a surprise is that,
despite this restriction, the turbulence still manages to transport
momentum efficiently across the separatrix.

\section{Discussion}\label{discussion}

The inclusion of the third dimension in numerical simulations of
magnetopause reconnection permits the development of strong turbulence
regardless of the presence of a guide field.  In both cases the strong
density gradient drives the development of a form of lower hybrid
drift instability along the magnetic separatrices downstream from the
X-line.  In the guide field case magnetic shear is not effective in
stabilizing the LHDI downstream of the X-line because the rotation of
the magnetic field occurs over a scale size of the order of the width
of the magnetic island while the gradient in the plasma density is
highly localized across the magnetic separatrix. At late time there is
a balance between the steepening of the gradient as the magnetic
island expands into the low density magnetosphere and the turbulence
associated with LHDI. Very near the X-line, unlike the anti-parallel
case, LHDI is stabilized.  Nevertheless, an electromagnetic
instability develops with a wavelength that greatly exceeds that due
to LHDI.  Although the turbulence in all instances contributes to
balancing the reconnection electric field, the overall reconnection
rate is essentially unaffected.

An important issue related to the impact of turbulence driven by the
LHDI at the magnetopause concerns the response of electrons to the
fluctuations. Because the turbulence is low frequency compared to the
electron gyrofrequency and because the $\boldsymbol{\nabla}B$ drift
velocities of electrons are typically small compared with the phase
speed of the wave, electrons are typically frozen-in to the
fluctuations. This frozen-in behavior has been documented with the
observations from the MMS mission \cite{graham19a}. Within linear
theory electrons are therefore non-resonant and cannot undergo
irreversible diffusion nor contribute to the average Ohm's law
describing large-scale reconnection. However, we have shown that the
LHDI-driven turbulence reaches large enough amplitude for the
electrons to undergo fluid-like turbulent diffusion
\cite{kleva84a}. In this regime the electrons experience a nonlinear
resonant interaction with the fluctuations. This turbulent diffusion
can drive transport of both the electron density $n$ and the
out-of-plane current density $nV_{eM}$.

Reconnection in asymmetric configurations can be stabilized by the
presence of diamagnetic drifts \cite{swisdak03a,swisdak10a,phan13a},
with complete stabilization occurring when the difference in $\beta =
8\pi P/B^2$ between the asymptotic plasmas exceeds $\approx 2
\tan{\theta/2}$, where $\theta$ is the shear angle between the
reconnecting fields.  In the configuration considered here, $\Delta
\beta \approx 2.5$ and $2\tan{(\theta/2)} \approx 4.8$.  Hence the
reconnection is not strongly affected by diamagnetic drifts, which is
in agreement with the reconnection rate of $\mathcal{O}(0.1)$ observed
for the both the two-dimensional and three-dimensional simulations.

An important question is whether real mass-ratio simulations (here
$m_i/m_e=100$) would give different results.  Even with a realistic
mass ratio, the LHDI will be strong in systems with scale lengths near
the ion Larmor radius, which is characteristic of the boundary layers
with strong $E_N$ at the magnetopause. The suppression of LHDI by
magnetic shear and finite $\beta$ is weaker in asymmetric reconnection
because the strongest density gradient and peak current $J_{eM}$,
which drive the instability, are on the magnetosphere side of the
X-line where $\beta$ is smaller.  Comparisons of simulations of
anti-parallel reconnection with $m_i/m_e=100$ and $m_i/m_e=400$ found
qualitative similarities, although the amplitude of the LHDI and its
extent in the $N$ direction were greater in the latter case
\cite{price17a}.  We anticipate similar results will hold for the
guide-field case along the separatrices away from the X-line.
However, the scaling near the X-line is less certain, particularly for
the inertial terms that make a significant contribution to the
generalized Ohm's law in Figure \ref{ohmplot_x}(b).

\appendix
\section{Averaged Generalized Ohm's Law}\label{appA}

In order to derive the various contributions arising from turbulent
fluctuations, begin with the momentum equation for the electron fluid
\begin{equation}\label{ohmeq}
en\mathbf{E} = -mn\frac{d\mathbf{v}}{dt} - \boldsymbol{\nabla
  \cdot}\mathbb{P} - en(\mathbf{v}\boldsymbol{\times}\mathbf{B})/c
\end{equation}
where $m$, $n$, $\mathbf{v}$, and $\mathbb{P}$ are the electron mass,
density, velocity, and pressure tensor and $d/dt$ represents the total
(convective) derivative.  Next, average over the $M$ direction and
decompose every quantity into a mean and fluctuating component, i.e.,
$n = \langle n \rangle + \delta n$. Note that products of quantities
produce two potentially non-zero terms, $\langle AB \rangle = \langle
A\rangle \langle B\rangle + \langle \delta A \delta B\rangle$ and
triple products (e.g., $n\mathbf{v}\boldsymbol{\times}\mathbf{B}$)
produce five, including one average of three fluctuating terms.
Previous applications of this approach \cite{price16a,price17a,le17a}
have combined the number density and fluid velocity in the final term
of equation \ref{ohmeq} into a single current density term
$\mathbf{J}$.  Here, in contrast, we separate $\mathbf{J}$ into its
constituent parts in the Lorentz force term -- although not the terms
proportional to the mass $m$ -- in order to explore the degree to
which electrons remain frozen to the magnetic field.  The final result
is
\begin{equation}\label{wholething}
\begin{split}
e\langle n\rangle \langle E_M \rangle = 
&\frac{e}{c}\Bigl(\langle n\rangle\langle v_L\rangle\langle B_N\rangle 
- \langle n\rangle\langle v_N\rangle\langle B_L\rangle\Bigr) \\
&-\frac{\partial}{\partial L} \langle P_{LM}\rangle -
\frac{\partial}{\partial N} \langle
P_{NM}\rangle\\ 
&+\frac{m}{e}\left(\frac{\partial}{\partial L}\langle
v_L\rangle\langle J_M\rangle 
+ \frac{\partial}{\partial N}\langle v_N\rangle
\langle J_M\rangle + \frac{\partial}{\partial t}\langle J_M\rangle\right)\\ 
&-e\langle\delta n\delta E_M\rangle \\
&+ \frac{m}{e}\left(\frac{\partial}{\partial L}\langle\delta
J_M\delta v_L\rangle + \frac{\partial}{\partial N}\langle\delta
J_M\delta v_N\rangle\right)\\ 
&+\frac{e}{c}\Bigl(\langle n\rangle\langle \delta v_L\delta B_N\rangle
-\langle n\rangle\langle \delta v_N\delta B_L\rangle \\
&\qquad+\langle B_N\rangle\langle \delta n\delta v_L\rangle 
-\langle B_L\rangle\langle \delta n\delta v_N\rangle \\
&\qquad+\langle v_L\rangle \langle\delta n\delta B_N\rangle 
-\langle v_N\rangle \langle \delta n\delta B_L\rangle\\
&\qquad+\langle \delta n\delta v_L\delta B_N\rangle
-\langle \delta n\delta v_N\delta B_L\rangle\Bigr)
\end{split}
\end{equation}

\begin{acknowledgments}
This work was supported by NASA grants NNX14AC78G, NNX16AG76G, and
80NSSC19K0396 and NSF grant PHY1805829. The simulations were carried
out at the National Energy Research Scientific Computing Center.  The
data used to perform the analysis and construct the figures for this
paper are preserved at the NERSC High Performance Storage System and
are available upon request.

\end{acknowledgments}

%


\end{document}